\documentclass[manuscript]{aastex}
\usepackage{amssymb,txfonts,mathrsfs,rotating}

\begin{document}

\title{Mass Segregation of Embedded Clusters in the Milky Way}

\author{Xinyue Er\altaffilmark{1,2} and Zhibo Jiang\altaffilmark{1}and Yanning Fu\altaffilmark{1}}
\affil{1 Purple Mountain Observatory, Chinese Academy of Sciences,
Nanjing 210008} \affil{2 Graduate School, Chinese Academy of
Sciences, Beijing 100039} \email{exy@pmo.ac.cn zbjiang@pmo.ac.cn
fyn@pmo.ac.cn}

\begin{abstract}
Embedded clusters are ideal laboratories to understand the early phase of the dynamical evolution of clusters as well as the massive star formation.
An interesting observational phenomenon is that some of the embedded clusters show mass segregation, i.e.,
the most massive stars are preferentially found near the cluster center.
In this paper, we develop a new approach to describe mass segregation.
Using this approach and the Two Micron All Sky Survey Point Source Catalog (2MASS PSC), we analyze eighteen embedded clusters in the Galaxy.
We find that eleven of them are mass-segregated and that the others are non-mass-segregated.
No inversely mass-segregated cluster is found.
\end{abstract}


\section{Introduction}
In recent years, the development of near infrared instruments has deepened our knowledge of embedded clusters in the Galaxy.
Some of the embedded clusters show mass segregation, i.e., the most massive stars are preferentially found near the cluster center.
This phenomenon has been observed in the Trapezium\citep{Hillenbrand97,Hillenbrand98},
NGC6611\citep{Bonatto06}) M17\citep{jiang02}, NGC1333\citep{Lada96}, NGC2244 and NGC6530\citep{chen07}.
More details on this topic can be found in the reviews (\citet{elmegreen00,Lada03}).

Mass segregation of embedded clusters can be dynamical.
\citet{McMillan2007} find that mass-segregated clusters can be quickly formed by merging several subclusters.
Simulations by \citet{Allison09mst} and \citet{yu2011}
confirm that the violent evolution of a cool and fractal cluster can give rise to mass segregation in a short timescale ($\sim$ 1Myr).

Mass segregation of embedded clusters can also be primordial.
According to Jeans theory, Jeans mass tends to be smaller, thus yielding less massive protostars,
due to higher density in the center of a molecular core than that in the outskirts,
whereas these protostars will accumulate gas and eventually evolve into massive stars more easily through competitive accretion\citep{larson82,ML96,bbcp97}.
In addition to the mechanism of competitive accretion, it is argued that the protostars are so rich in the cluster center that
they can merge into the massive stars\citep{bbz98,bb05}.

Moreover, mass segregation of embedded clusters may not be ``true".
For instance, \citet{Ascenso09} argue that it might be an observational bias in some cases.
\citet{er09} argue that it might be a temporary aggregation resulting from the random motions of massive stars.

It can be seen that studying mass segregation of embedded clusters will help us understanding the early dynamical evolution of clusters
and the massive star formation.
However, so far it has not been clear whether or not mass segregation is a common phenomenon associated with embedded clusters.
Consequently, we analyze the mass segregation of eighteen clusters in our Galaxy in this paper.
In section 2, we describe our approach based on a new index---$\mathscr{R}$.
In section 3, with realistic clusters, we show the validity of the approach.
In section 4, we analyze the status of the mass segregation of eighteen clusters.
In section 5, we discuss the implications of our results.
In section 6, a summary is given.

\section{Description of Mass Segregation}
\subsection{A Brief Review}

\citet{Hillenbrand97} uses the variation of the ratio of massive stars to low-mass stars in different regions to probe mass segregation.
Mass segregation can also be reflected in the variations of mean stellar mass\citep{Hillenbrand98}, mass function, and
luminosity function\citep{Hunter95,Brandl96,Vazquez96,Fischer98,de02,Kerber06}.
Nevertheless, as pointed out by \citet{Gouliermis04} and \citet{MU05}, 
one should note the uncertainty caused by the determination of the slope of power laws.

From another viewpoint, the distribution of massive stars is more concentrated than that of low-mass stars in a mass-segregated cluster.
This will lead to that the half-number radius of the massive stars is smaller than that of low-mass stars\citep{zhao06}.
Also, the surface number density profiles of massive stars and low-mass stars are different\citep{lada91}.
If profiles are characterized by power-law, the indices are different\citep{Sagar88,Kontizas98},
if profiles are characterized by the King model\citep{king62,king66}, the core radii are different\citep{Nurn02}.
Moreover, the profiles can be characterized by different models.
For M17, \citet{jiang02} find an exponential radial decline for massive stars and a power-law radial decline for low-mass stars.
Sometimes the profiles are transformed into cumulative forms in which their differences are checked by the Kolmogorov-Smirnov test\citep{zhao06,chen07}.
Actually, the Kolmogorov-Smirnov test can be directly applied to the distributions of massive stars and low-mass stars\citep{Hillenbrand98,Raboud98}.

Recently, \citet{Allison09mst} introduce $\Lambda$,
the ratio of the length of the minimum spanning tree of massive stars to that of low-mass stars, to characterize mass segregation.
The advantage of this index is that it does not rely on the defining of cluster center.
In the present work, we develop a new approach to describe mass segregation.

\subsection{A New Index}
We define the new index as $\mathscr{R} = \frac{\overline{L}_{\rm part}}{\overline{L}_{\rm all}}$,
where $\overline{L}_{\rm part}$ is the mean mutual distance of a special class of stars, and $\overline{L}_{\rm all}$ that of all stars.
If $\mathscr{R}$ $<$ 1, the distribution of the special class of stars is more concentrated.
The smaller $\mathscr{R}$ is, the more pronounced the concentration.
When the special class of stars refers to top massive stars, $\mathscr{R}$ becomes an index of mass segregation.

Note that the deviation of $\mathscr{R}$ from unity does not necessarily mean mass segregation, for it can be merely a consequence of fluctuation.
In order to cope with the fluctuation, numerical tests have been performed to obtain a reasonable threshold of $\mathscr{R}$.
We generate 100 cluster samples, each consisting of 1000 stars with different masses\citep{Cartwright04}.
The stars are mass-independently distributed following a surface number density profile in the form of $\rho \propto r^{-1}$, where $r$ is the radial distance.
Provided that the number of top massive stars ($N_{\rm top}$) is fixed,
$\mathscr{R}$ can be well fitted by a Gaussian distribution within the confidence interval of 3$\sigma$,
where $\sigma$ is the standard deviation (See Figure 1).
Following the definition of Gaussian distribution, the samples far smaller than unity can be regarded as mass segregation.
In other words, we can obtain the threshold of $\mathscr{R}$ from $\sigma$.
Further tests suggest that $\sigma$ is related to $N_{\rm top}$, since the width of the Gaussian curve becomes narrower as $N_{\rm top}$ increases from 10 to 30.
Figure 2 shows $\sigma$ is a function of $N_{\rm top}$.
When $N_{\rm top}$ is small, $\sigma$ is extremely large and declines rapidly with the increase of $N_{\rm top}$.
For larger $N_{\rm top}$, the change of $\sigma$ becomes smaller.
This suggests that the dependence of $\sigma$ on $N_{\rm top}$ should be taken into account.
Indeed, this also illustrates that mass segregations deduced from only a few stars are inherently uncertain, as \citet{Lada03} argued.
We also generate two other sets of cluster samples in which the numbers of cluster members ($N_{\rm all}$) are 500 and 2000.
Their dependences of $\sigma$ on $N_{\rm top}$ are obtained and presented in Figure 2.
It can be seen that effect of $N_{\rm all}$ is much weaker than that of $N_{\rm top}$.
Thus, we do not consider its effect in this paper.
The number density profiles of realistic clusters are generally different.
They can be roughly represented by the form of $\rho \propto r^{-\alpha}$.
Figure 3 shows that $\sigma$ grows with an increasing $\alpha$.
This suggests that the effect on $\sigma$ due to the profiles should be considered.

For a given cluster of 1000 stars, we select ten stars as a set and calculate its $\mathscr{R}$ and $\Lambda$, respectively.
In order to study their relations in different environments, we select many sets in which stars are distributed at different degrees of concentration.
Figure 4 shows that they have a good correlation, which indicates that $\mathscr{R}$ is another choice of describing mass segregation.
It is worth mentioning that the consuming time of $\mathscr{R}$ is $\propto$ $N$, while that of $\Lambda$ is $\propto$ $N^2$.

\subsection{$\mathscr{R}-N_{\rm top}$ Plot}
Obviously, the $\mathscr{R}$ value of a cluster depends on the chosen $N_{\rm top}$.
So a $\mathscr{R}-N_{\rm top}$ plot is introduced to describe the status of the mass segregation of a cluster.
Figure 5 shows the $\mathscr{R}-N_{\rm top}$ plots of four typical artificial clusters.

Panel (a) shows the case of a non-mass-segregated cluster.
The stars in the cluster are mass-independently distributed.
Although some of the $\mathscr{R}$ are lower than unity, few of them are lower than unity with 1$\sigma$ confidence.
This kind of deviation of $\mathscr{R}$ from unity can be viewed as a fluctuation.
Panel (b) shows the case of general mass segregation.
In this cluster we place the top five percent of stars inside the half number radius of the cluster.
We find nearly all the values of $\mathscr{R}$ are lower than unity and most of them are lower than unity with 1$\sigma$ confidence.
Panel (c) shows the case of dynamical mass segregation.
In this cluster, the radial distance of each star is strictly related to its mass,
with the most massive stars located innermost and the lowest-mass stars outmost.
As is shown, $\mathscr{R}$ have a smooth increase in a wide mass range.
Panel (d) shows the case of primordial mass segregation.
In this cluster, the top five massive stars are in the center region and the other stars are mass-independently distributed.
One may find an abrupt increment of $\mathscr{R}$ at $N_{\rm top}=5$.
Although we only rearrange the top five massive stars, the effect seems to exist until $N_{\rm top} \sim 100$.
This is because these five massive stars are located in the very center of the cluster,
and $\mathscr{R}$ at $N_{\rm top}=100$ contains all the position information from $N_{\rm top}= 2$ to $100$.

\subsection{Definition of Mass segregation for a Cluster}
The primary goal of this paper is to study whether or not mass segregation is a common phenomenon for embedded clusters.
Therefore we set a definition to classify clusters into two categories, i.e., with or without mass segregation,
ignoring the details of $\mathscr{R}-N_{\rm top}$ plot.
Considering that the dispersion of $\mathscr{R}$ is dramatically large for small $N_{\rm top}$,
we restrict ourselves to $5 \leqslant N_{\rm top} \leqslant N_{\rm all}$.
In this range we try to find the largest number of $N_{\rm top}$, denoted as $N_x$,
such that the values of  $\mathscr{R}$ from $N_{\rm top}$ = 5 to $N_x$ satisfy:

1. They are all lower than unity.

2. Half of them are lower than $1-s\times \sigma$, where s is called the level of mass segregation. In this paper, we choose s = 1 or 3.

If $N_x$ exists for a given s, we consider the cluster is level-s mass-segregated in the range from $N_{\rm top}$ = 5 to $N_x$;
if $N_x$ does not exist, we consider the cluster is non-mass-segregated.
We believe that this quantitative definition can distinguish between fluctuation and real mass segregation.

\section{Test of Validity of Our Approach}
\subsection{Sampling a Cluster}
The positional and photometric data of the clusters are extracted from 2MASS PSC in Ks band.
In order to guarantee the reliability, the following data are excluded from consideration.

1. ``Kmag $>$ 14.3 mag", for 14.3 mag is the limiting magnitude of Ks band.
Most of the abandoned stars are due to this reason.

2. ``Qflg = U", which means the catalog only gives the upper limit on magnitude.

3. ``use = 0", which means the source is an apparition.

4. ``Xflg = 0" and ``extkey is null", which means the source is an extragalactic source.

4. ``Aflg = 1", which means the source is associated with a known solar system object.

5. ``Qflg = X", which means there is no valid brightness, although a detection is found.

The top panel of Figure 6 shows the surface density map of Orion Nebula Cluster (ONC) in the 30' $\times$ 30' field.
Assuming that the most populated area is the cluster center, we construct its radial density profile in the bottom panel of Figure 6.
The uniform model $\rho(r) = \frac{C_0}{r^{C_1}}+C_2$ is used as a fitting model of the profile,
where $C_0$ is a fitting parameter, $C_1$ is the index of the profile, and $C_2$ stands for the surface number density of background stars.
We truncate the cluster at the radius where $\rho(r) = 3C_2$.
There are cases in which the best fitting value of $C_2$ is negative.
To avoid this and to be consistent in processing all sample clusters, we do not adjust $C_2$ in the fitting.
Instead, we fix its value roughly as the mean density of the background.
All the clusters considered in this paper are truncated in the same way as above.

Identification of members of a cluster is considerably difficult.
\citet{jayking03,chen07,pang2010} use proper motion to identify the memberships.
However, this method needs a long term observation that spans many years.
\citet{Soares02,Bonatto03,Bonatto05} use color-magnitude and color-color diagrams to identify the memberships.
However, this method still contains uncertainties from the photometry and the evolutionary track.
Because of the facts that the associated clouds give rise to the severe extinction of embedded clusters and that our cluster samples are all in 2kpc,
we estimate that the background and foreground stars within the truncated radius of a cluster are less than 10\%.
That is, the effect of contamination is statistically insignificant.
As a result, we regard that all the stars in the truncated radius are cluster members.

\subsection{Case of ONC}
ONC is the most famous star formation region with a mean age less than 1Myr (Hillenbrand \citep{Hillenbrand97}.
The cluster shows apparent mass segregation\citep{Hillenbrand97,Hillenbrand98}.
We take it as an example to show the validity of our approach.
Figure 7 shows the $\mathscr{R}-N_{\rm top}$ plot of ONC.
Generally speaking, $\mathscr{R}$ has an increasing trend with $N_{\rm top}$, although the trend is non-monotonic.
Following our definition, ONC is mass-segregated.
In fact, the mass segregation is so pronounced that ONC can be viewed as a level-3 mass-segregated cluster.

\citet{Allison09mst} also consider that ONC is mass-segregated,
whereas they argue that ONC is only mass-segregated for the top ten massive stars.
This inconsistency results from the different cluster members are used.
Specifically, our cluster's center is in agreement with theirs, but our cluster's extent is about half of theirs.
Moreover, significantly more dim stars are detected by 2MASS in this region.
Notice that \citet{Allison09mst} argues their data set may be lack of low-mass stars, since they only use the stars that are provided with masses.

\subsection{Case of Mon R2}
Mon R2 is another embedded cluster close to us.
As is shown in Figure 8, its $\mathscr{R}-N_{\rm top}$ plot is quite different from that of ONC.
$\mathscr{R}$ is a little larger than unity at the beginning, then falls until $N_{\rm top}$ $\sim$ 60, and then rises again.
$\mathscr{R}$ is lower than $1 - \sigma$ from $N_{\rm top}$ $\sim$ $30$ to $100$.
These facts indicate that the distribution of the most massive stars is more scattered than that of all stars,
but quite a few intermediate massive stars are distributed in the center region.
Following our definition, Mon R2 is non-mass-segregated.
\citet{Carpenter97} also consider that Mon R2 does not present compelling evidence of mass segregation.

\section{Results}
Our embedded cluster samples come from the catalog of \citet{Lada03}.
However, using the method described in Section 3, we only identify eighteen clusters from their catalog.
Notice that we require the cluster's density is three times more than the background.
So the clusters that have a high contamination surrounding the cluster are failed to be identified.
Likewise, some clusters are excluded from consideration due to their scarcity of cluster members.
\citet{Lada03} argue that 35 stars can make the cluster survive evaporation during its lifetime,
so the clusters in their catalog have more than 35 cluster members.
But the short exposure time of 2MASS PSC and different adopted cluster radii cause some of the clusters have less than 35 members in our analysis.
Considering that a sufficient number of stars is also necessary for statistical significance,
these clusters are not taken into account in this paper.
It is worth mentioning that, in order to enrich our cluster samples,
we have tried other embedded cluster catalogues\citep{Dutra01,Dutra03}, but no new sample is found.

\begin{table}
\caption{The catalogue of 18 embedded clusters}
\begin{tabular}{c c c c c c c c cr cr}
\hline \hline
EC& name & RA Dec & distance & radius & $N_{\rm all}$ &  mass segregation &  mass segregation \\
  &      &(J2000) & (pc)     &  (pc)  &               &     status        &     range($N_x$)    \\
\hline
  1&             NGC2071&   05 47 08.0  +00 20 49&      400& 0.26&    39&   N& -- & \\
  2&         LKHalpha234&   21 43 00.0  +66 06 59&      1000&  0.43&   51&   N& -- &  \\
  3&               Gem4&   06 08 43.0  +21 31 19&      1500&  0.53&   56&    N& -- &  \\
  4&               NGC1333&   03 29 04.0  +31 21 24&      318&  0.40&   77&  N& -- &   \\
  5&             W3IRS5&   02 25 39.0  +62 06 22&      2400&  1.35&   157&    N& -- & \\
  6&          LKHalpha101&   04 30 12.0  +35 16 55&      800& 0.90&   157&  N&  -- &  \\
  7&               Mon R2&    06 07 45.0  $-$06 23 29&      800& 0.97&    306&  N&  -- &  \\
  8&              L1654&    06 59 44.0  $-$07 46 59&      1100& 0.32&   44&   Y& 5& \\
 9&            NGC2244&   06 34 13.0  +04 26 43&     1600&  0.64&   44&  Y& 11 &   \\
 10&              S235B&   05 40 55.0  +35 40 55&      1800& 1.28&   141&  Y& 15 &   \\
 11&         AFGL5157&   05 37 46.0  +31 59 54&      1800& 0.56&    37&  Y& 25 &   \\
 12&            IC5146&    21 53 27.0  +47 15 31&      1200& 0.48&    62&   Y& 41 &   \\
 13&                IC348&    03 44 36.0  +32 08 46&      320&  0.23&      76&  Y& 63 &   \\
 14&            GGD12-15&   06 10 49.0  $-$06 12 24&      800& 0.58&    79&   Y& 59 &  \\
 15&            NGC2024&   05 41 45.0  $-$01 55 00&      400& 0.57&   334&   Y& 71 &  \\
 16&                CepA&   22 56 21.0  +62 02 27&      700&  0.27&   77&  Y&  66 &  \\
 17&               RCW38&   08 59 03.0  $-$47 30 12&      1700& 0.73&   202&  Y&  202 &  \\
 18&                 ONC&   05 35 19.0  $-$05 22 44&      450&  0.96&  1216&  Y& 682 &  \\
\hline
\label{tab:table1}
\end{tabular}
\end{table}

The $\mathscr{R}-N_{\rm top}$ plots of the eighteen clusters are presented in Figure 9.
Table 1 lists their names, locations, distances, radii, numbers of members, status of mass segregation, and mass segregation range.
For NGC2071, LKHalpha234, Gem4, NGC1333, W3IRS5, and LKHalpha101, all the $\mathscr{R}$ are close to unity, so they are non-mass-segregated.
Mon R2 also belongs to this category.
L1654, NGC2244, S235B, AFGL5157, IC5146, IC348, and GGD12-15 are level-1 mass-segregated at a certain range (See the details in Table 1).
Mass segregation of NGC2024, CepA, and RCW38 are rather pronounced.
They are all level-3 mass-segregated at a certain range (from $N_{\rm top}$=5 to 15, 65, and 202).
ONC also belongs to this category.

In conclusion, according to our definition, eleven clusters are level-1 mass-segregated,
among which four clusters (NGC2024, CepA, RCW38, and ONC) are level-3 mass-segregated.
The other clusters are non-mass-segregated.
No cluster is found with convincing evidence for inverse mass segregation.

\section{Discussions}

\subsection{Variation of Parameter in Data Processing}
The limiting magnitude of unconfused regions of 2MASS is 14.3 mag for Ks band.
However, for the crowded cluster center, the limiting magnitude might be less than 14.3 mag.
Therefore, for the eighteen clusters, we reduce the limiting magnitude from 14.3 mag to 13.3 mag to study the effect of stellar crowding.
In this case, four clusters have less than 35 members, so they are removed from the test.
Some $\mathscr{R}-N_{\rm top}$ plots of new clusters are presented in Figure 10.
We find the judgements of the mass segregation of the remaining fourteen clusters hold.
This means that the effect of stellar crowding should not affect our results.

Observationally, the determination of radius always has some uncertainties.
To study the effect, we make the radius smaller than the adopted value in Section 4, assuming the uncertainty is 20\%.
In this case, sixteen clusters have more than 35 members.
Some of the new $\mathscr{R}-N_{\rm top}$ plots are shown in Figure 11.
Again, although the details are changed, the judgements of mass segregation hold.
This means that the uncertainty of the radius is not likely to affect our results.

\subsection{Occurrence of Mass-segregated Clusters and Implications}
For the artificial clusters in which the stars are mass-independently distributed,
we find 1\% of them are level-3 mass-segregated, 27\% of them are level-1 mass-segregated, and some of them show inverse mass segregation.
We also randomly choose some control regions in the whole sky and assume all the stars inside form a ``cluster".
The results are quite similar to that of artificial clusters.
These facts suggest that mass segregation observed in embedded clusters cannot always be an accidental phenomenon, especially for level-3 mass segregation.
We consider that level-3 mass segregation must be imprinted by the early dynamical evolution or the star formation of embedded clusters.

Another impressive thing is the deficiency of inversely mass-segregated clusters in observation.
This might be caused by the rapid dynamical evolution of inversely mass-segregated clusters.
In other words, inverse mass segregation is not a stable status for a cluster, which makes it hardly observed.
Note that \citet{Vesperini2009} show that initial mass segregation plays an important role in cluster survival.
Also note that not every embedded cluster can survive from the state of molecular cloud to open cluster\citep{Adams2001,Lada03}.
From this perspective, as more observations of embedded clusters, inversely mass-segregated clusters might be found.

\subsection{What Kind of Embedded Cluster is Likely to be Mass-segregated?}
We find some clusters are non-mass-segregated.
This is not likely to be caused by the inappropriate process bias,
because some clues for non-mass-segregated clusters are found in the survey of literature\citep{lada91,Carpenter97,Herbig02}.
So we believe the non-mass-segregated clusters do exist.
Then what kind of embedded cluster is likely to be mass-segregated?

We examine the relationship between the existence of mass segregation and the radius of embedded clusters.
No correlation is found, which is consistent with \citet{Hasan2011}.
Besides, we find the number of cluster members appears to be related to mass segregation.
For these eighteen clusters, the average number of members is 175.
That of the mass-segregated clusters is 210 and that of the level-3 mass-segregated clusters is 457.
So it seems that the richer clusters tend to be mass-segregated.

\subsection{Origin of Mass Segregation}
\citet{bd98} argue that embedded clusters are too young to show dynamical mass segregation, while some works show that mass segregation can be achieved by rapid dynamical evolution\citep{McMillan2007,Allison09mst,yu2011}.
This suggests that we cannot simply deduce the origin of mass segregation of embedded clusters from their age.
Then how to infer its origin?
Velocity---mass dependence of cluster member can provide useful information.
Specifically, if a cluster does not show this dependence, the cluster is not dynamical relax, mass segregation in the non-relaxed cluster should be primordial.
By this idea, \citet{chen07} and \citet{pang2010} verify that the mass segregations in NGC2244, NGC6530, and NGC3603 are primordial.

We find that the shape of $\mathscr{R}-N_{\rm top}$ plot can be another method to deduce the origin of mass segregation.
$\mathscr{R}-N_{\rm top}$ plot clearly shows two different kinds of mass segregation in observation.
Mass segregation can exist in a rather large mass range, such as CepA, RCW38, and ONC.
In this case, the clusters are likely to be dynamically relaxed, so it is likely to be dynamical mass segregation.
On the other hand, mass segregation can only exist in the high-mass end of a cluster, such as NGC2024.
This kind of mass segregation seems to suggest that the top massive stars form by special mechanism.
So it is likely to be primordial mass segregation.

\section{Conclusions}
In this paper, we introduce a new approach, $\mathscr{R}-N_{\rm top}$ plot,
to describe the mass segregation of clusters, and then apply it to eighteen embedded clusters in our Galaxy.
The main points of this work are summarized as follows:

1. Eleven of the eighteen embedded clusters are mass-segregated, seven clusters are non-mass-segregated,
no inversely mass-segregated cluster is found.
That is, mass segregation is not a common phenomenon associated with embedded clusters.

2. The shape of $\mathscr{R}-N_{\rm top}$ plots reveals that there are two kinds of mass segregation,
which can be hints of the origin of mass segregation.
For dynamical mass-segregated cluster, its $\mathscr{R}$ should be lower than unity in a large range.
For primordial mass-segregated cluster, its $\mathscr{R}$ should be only lower than unity in the high-mass end.

3. We find that the richer clusters tend to present mass segregation.

4. Absence of inversely mass-segregated cluster suggests that the distribution of stars in embedded clusters is not totally mass-independent.

\begin{acknowledgements}
We are grateful to Richard de Grijs, M.B.N. Kouwenhoven, and Christoph Olczak for their discussions and suggestions.
This publication makes use of data products from the Two Micron All Sky Survey, which is a joint project of the University of Massachusetts and the Infrared
Processing and Analysis Center/California Institute of Technology, funded by the National Aeronautics and Space
Administration and the National Science Foundation.
This work is supported by the Chinese National Science Foundation (NSFC) 10873037, 10921063, 10733030, and 10833001.
\end{acknowledgements}

\begin{figure}
\centering
 \includegraphics[width=0.5\textwidth]{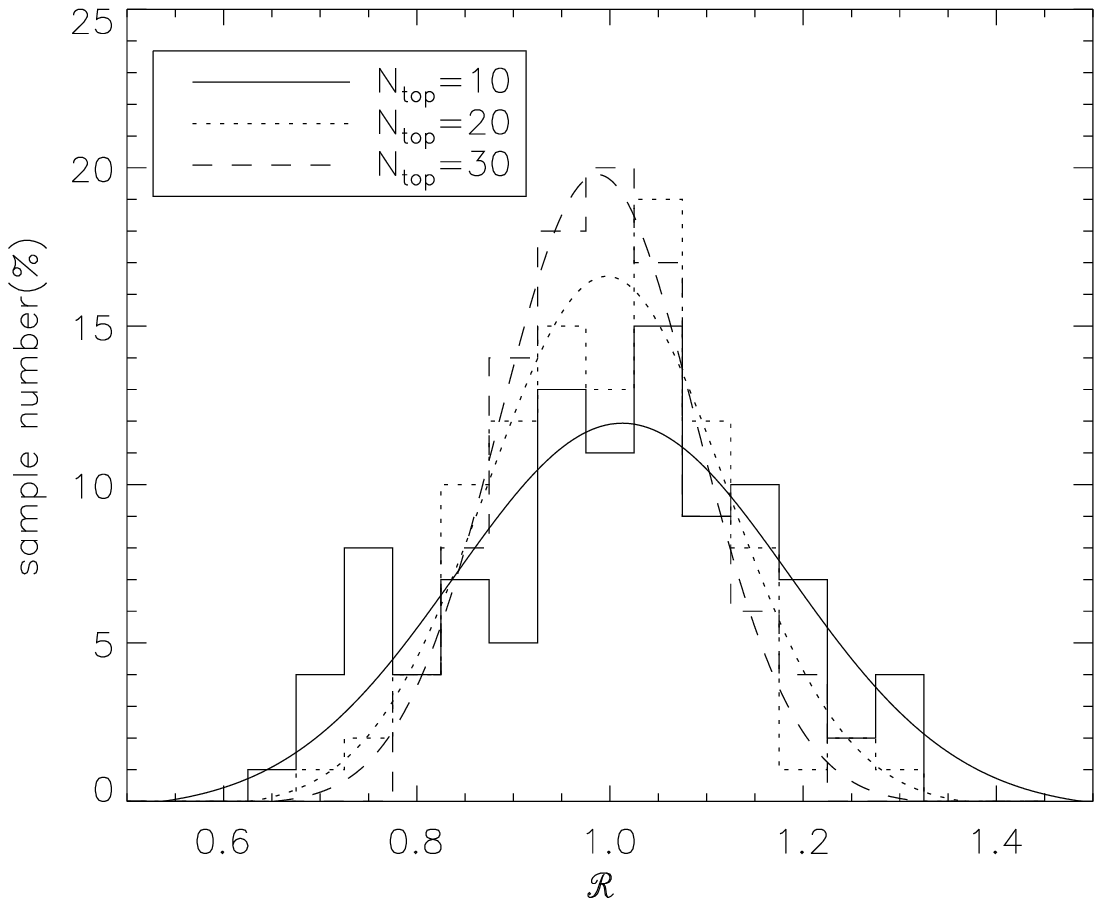}
 \caption{Histograms of $\mathscr{R}$ of 100 artificial cluster samples. We fit the values of $\mathscr{R}$ by Gaussian distribution.}
 \label{fig:fig1}
\end{figure}

\begin{figure}
\centering
 \includegraphics[width=0.5\textwidth]{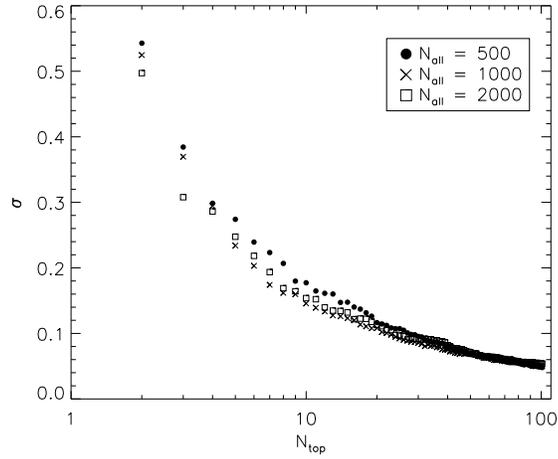}
 \caption{The dependences of $\sigma$ on $N_{\rm top}$ and $N_{\rm all}$.}
 \label{fig:fig2}
\end{figure}

\begin{figure}
\centering
\includegraphics[width=0.5\textwidth]{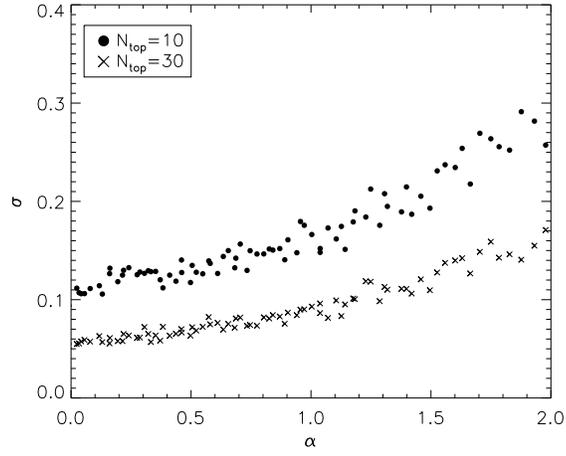}
\caption{The dependence of $\sigma$ on $\alpha$.}
\label{fig:fig3}
\end{figure}

\begin{figure}
\centering
\includegraphics[width=0.5\textwidth]{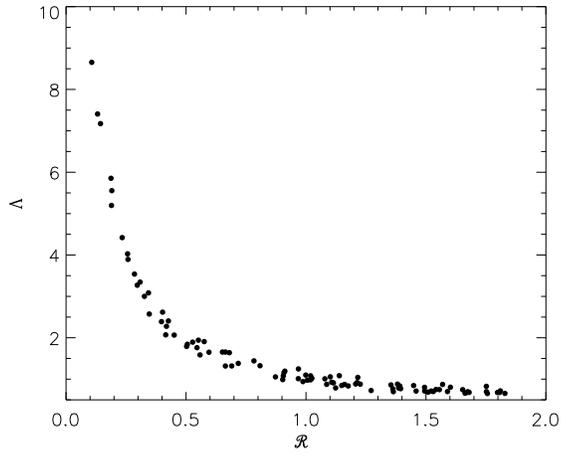}
\caption{Comparison between $\mathscr{R}$ and $\Lambda$.
 }\label{fig:fig4}
\end{figure}

\begin{figure}
\centering
\includegraphics[width=0.5\textwidth]{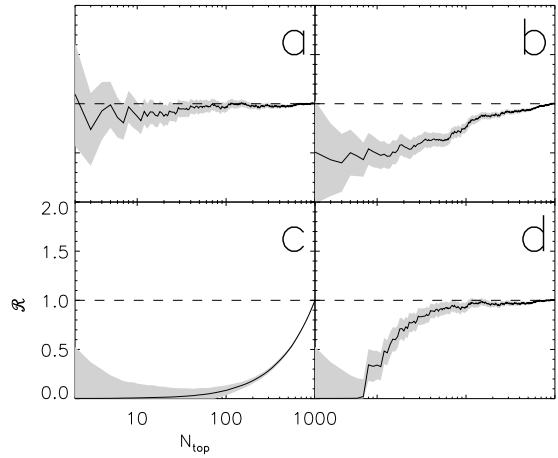} \caption{
$\mathscr{R}-N_{\rm top}$ plot of four typical artificial clusters.
The gray shaded band shows 1 $\sigma$ level confidence region of mass segregation.
Panel (a) is a non-mass-segregated cluster.
Panel (b) is a general mass-segregated cluster.
Panel (c) is the case of dynamical mass segregation.
Panel (d) is the case of primordial mass segregation.}
 \label{fig:fig5}
\end{figure}

\begin{figure}
\centering
\includegraphics[width=0.5\textwidth]{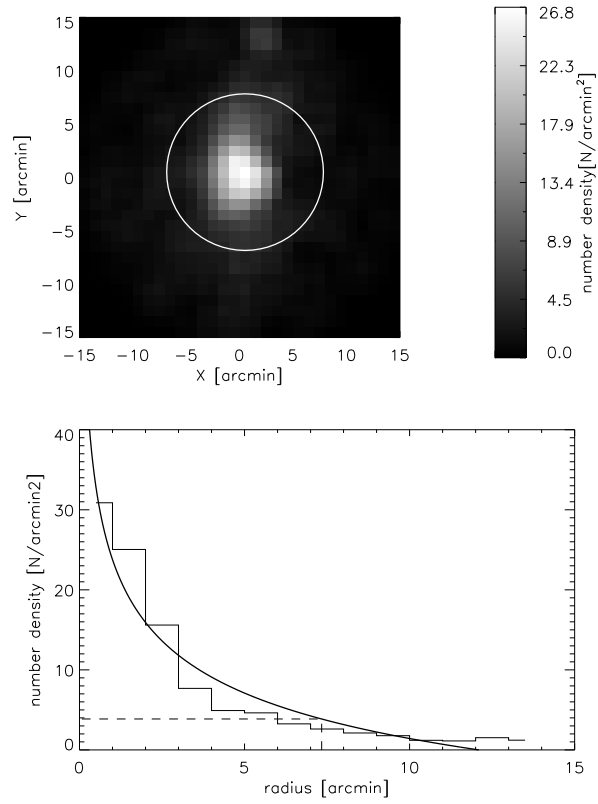}\caption{
Surface density map of ONC(top) and its surface number density profile(bottom).
The circle(top) and dash line(bottom) show the radius we determine.}
\label{fig:fig6}
\end{figure}

\begin{figure}
\centering
\includegraphics[width=0.5\textwidth]{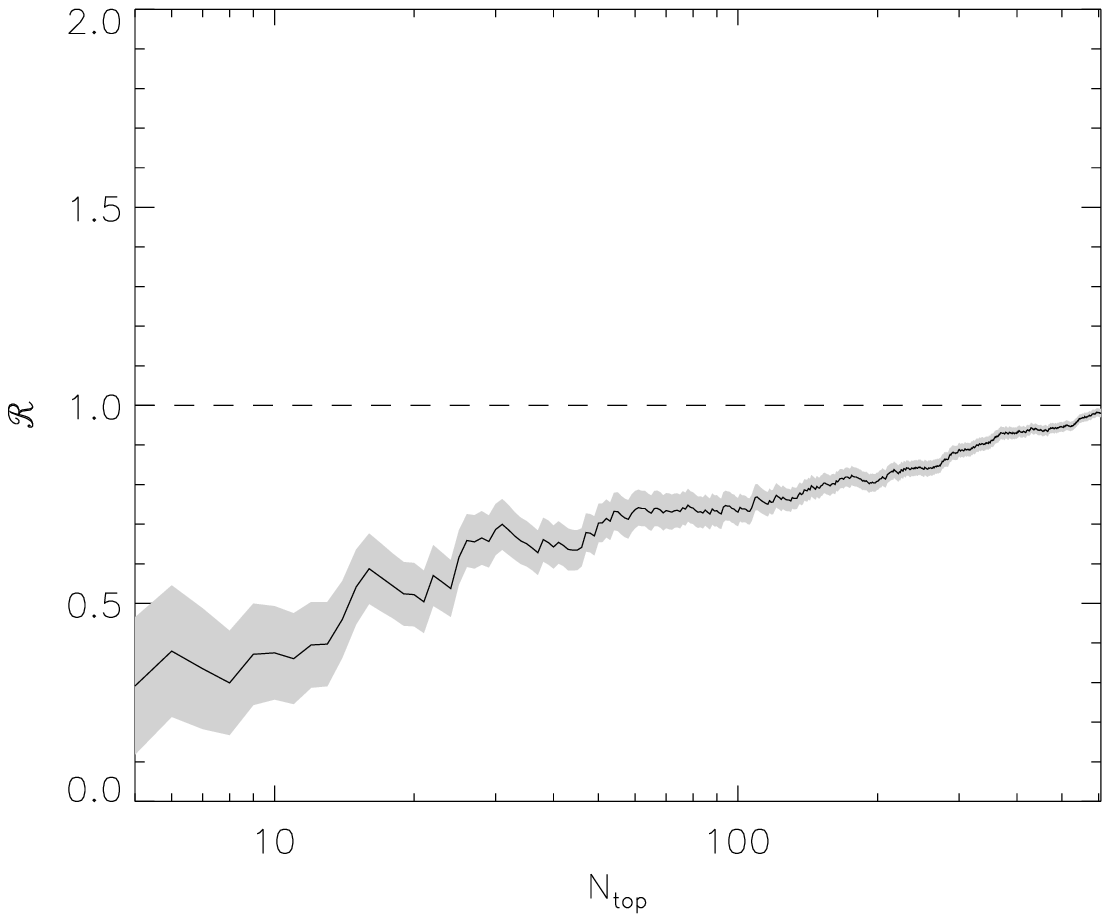} \caption{The $\mathscr{R}-N_{\rm top}$ plot of ONC.
Symbols denote the same as in Figure 5.}
\label{fig:fig7}
\end{figure}

\begin{figure}
\centering
\includegraphics[width=0.5\textwidth]{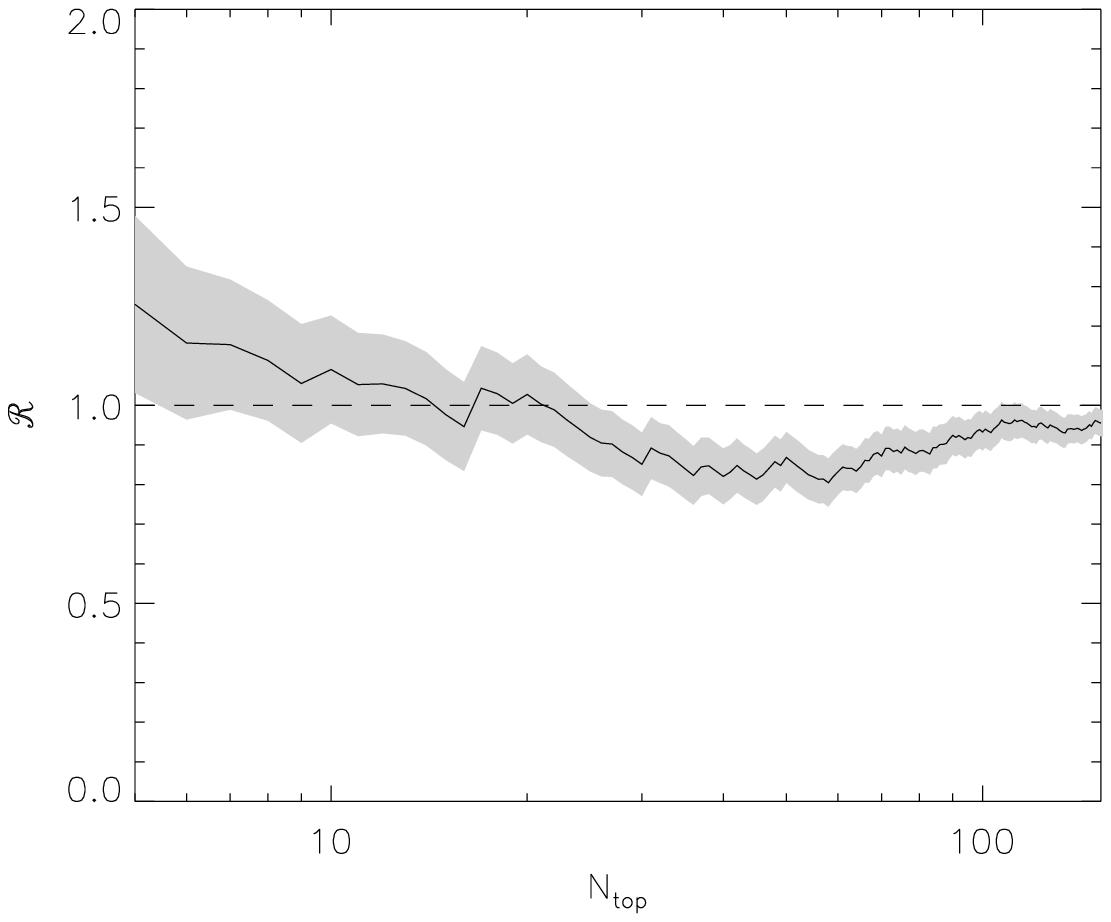}
\caption{The $\mathscr{R}-N_{\rm top}$ plot of Mon R2.
Symbols denote the same as in Figure 5.}
\label{fig:fig8}
\end{figure}

\begin{figure}
\centering
\includegraphics[width=0.5\textwidth]{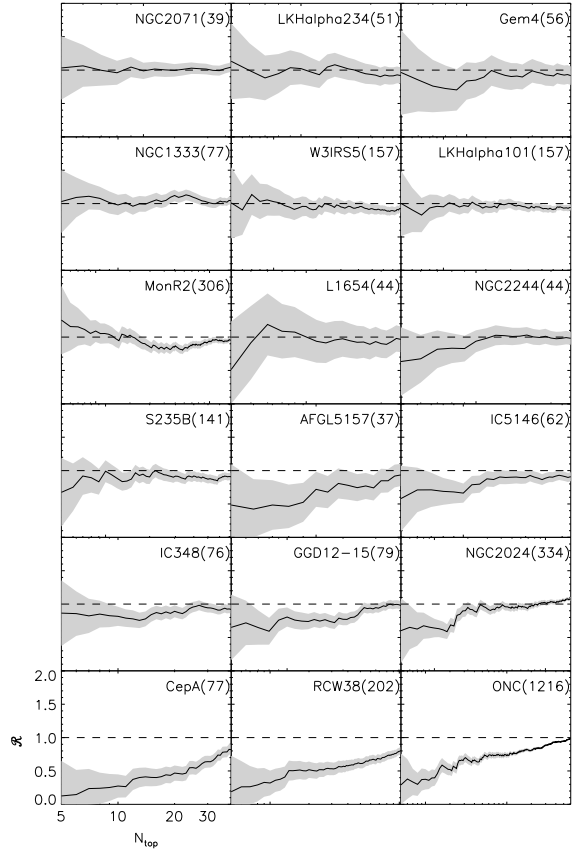} \caption{The $\mathscr{R}-N_{\rm top}$ plots of eighteen embedded clusters.
The cluster's name and number of stars are marked in the top right corner of each panel.
The gray shaded band shows 1 $\sigma$ level confidence region of mass segregation.}
 \label{fig:fig9}
\end{figure}

\begin{figure}
\centering
    \includegraphics[width=0.5\textwidth]{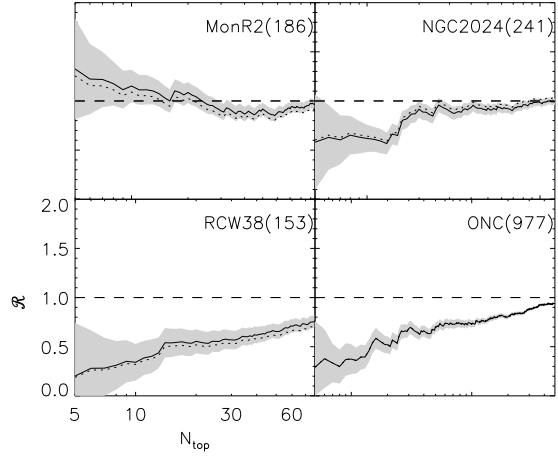}
    \caption{The $\mathscr{R}-N_{\rm top}$ plot of four clusters. Symbols denote the same as in Figure 9.
    The limiting magnitude is set to be 13.3 mag in this test to study the effect of stellar crowding.
    The dotted line represents the original values.}
    \label{fig:fig10}
\end{figure}

\begin{figure}
\centering
    \includegraphics[width=0.5\textwidth]{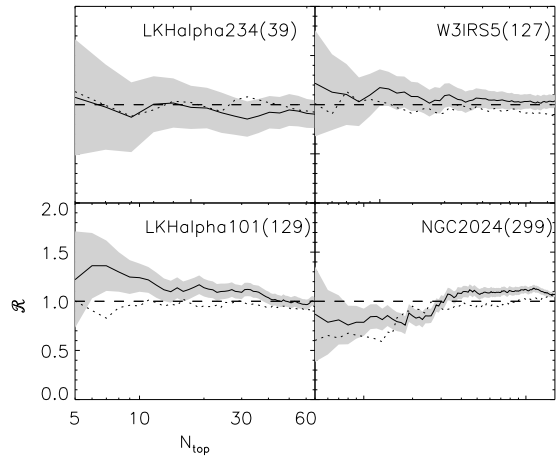}
    \caption{The $\mathscr{R}-N_{\rm top}$ plot of four clusters. Symbols denote the same as in Figure 10.
    In this test, the cluster radii shrinks to study the effect of the uncertainty of the radius.}
    \label{fig:fig11}
\end{figure}

\end{document}